\documentclass[10pt]{iopart}

\usepackage{cite}   

\begin{document}

\newcommand{\ol}[1]{\overline{#1}}

\title[The Efroimsky formalism adapted to high-frequency perturbations]{The Efroimsky formalism for weak gravitational waves adapted to high-frequency perturbations}

\author{O Sv\'\i tek\dag\  and J Podolsk\'y\ddag}

\address{Institute of Theoretical Physics, Faculty of Mathematics and Physics, Charles University in Prague, 
V~Hole\v{s}ovi\v{c}k\'ach 2, 180~00 Praha 8, Czech Republic }
\eads{\dag\ \mailto{ota@matfyz.cz}, \ddag\ \mailto{podolsky@mbox.troja.mff.cuni.cz}}

\begin{abstract}  
The Efroimsky  perturbation scheme for consistent treatment of gravitational waves and their influence on the background is summarized and compared with classical Isaacson's high-frequency approach. We  demonstrate that the Efroimsky method in its present form is not compatible with the Isaacson limit of high-frequency gravitational waves, and we propose its natural generalization to resolve this drawback.
\end{abstract}

\submitto{\CQG}

\maketitle

\section{Introduction}

Some time ago Efroimsky introduced and developed new formalism for a consistent treatment of weak gravitational waves \cite{efroimsky,efroimsky2}. This interesting mathematical framework is remarkable mainly due to the possibility to ascribe stress-energy tensor even to {\it low-frequency} gravitational waves influencing the background, which is in contrast to standard linearization approach where the background is kept fixed. This is achieved by introducing a natural low-frequency cut-off, employing three different metrics (the premetric, the complete physical metric, and the average metric) and careful analysis of their mutual relations.

On the other hand, in a now classic paper \cite{isaac} Isaacson (inspired by previous works \cite{wheeler,bh}) presented a perturbation method which can be used for studies of {\it high-frequency} gravitational waves. Such waves also influence the cosmological background in which they propagate. Isaacson's work stimulated further contributions in which his method was re-formulated using various formalisms, and explicitly applied to particular spacetimes, see e.g. \cite{choquet,maccallum-taub,araujo,araujo2,elster,burnett,taub,taub2,hogan,podolsky-svitek}.

In our present work we first briefly summarize and compare the two above mentioned perturbation schemes. In particular, it is shown that the Efroimsky method is not consistent if high-frequency gravitational waves are considered. Next (in section~3), we propose a possible modification of the Efroimsky formalism which may resolve this drawback.

\section{The formalism}

Efroimsky's approach \cite{efroimsky,efroimsky2} is based on introducing {\it three different} smooth, non-degenerate, symmetric  metrics on a differentiable manifold $M$, namely:
\begin{enumerate}
\item
$\gamma_{\mu\nu}\quad -$
the ``premetric'': vacuum metric corresponding to initial pure background without gravitational waves,
\item
$g_{\mu\nu}\quad -$
the ``physical metric'': full vacuum metric which describes both the background and the waves,
\item
$q_{\mu\nu}\quad -$
the ``average metric'': non-vacuum metric representing the background plus its perturbations with wavelength greater than $L$. In fact, it is the averaged full metric $g_{\mu\nu}$, where the cut-off value $L$ depends on observer's experimental abilities. Since no detector can measure gravitational waves of arbitrarily long  wavelengths, the existence of such low-frequency cut-off is a natural assumption.
\end{enumerate}
One motivation for using these three distinct metrics is to resolve a (slight) discrepancy in standard linearization approach which considers only the metrics $\gamma_{\mu\nu},$ $g_{\mu\nu},$ and decomposition $g_{\mu\nu}=\gamma_{\mu\nu}+h_{\mu\nu}\ ,$ where $h_{\mu\nu}$ is a small perturbation. The contravariant components obtained as an inverse of $g_{\mu\nu}$ are 
$g^{\mu\nu}=\gamma^{\mu\nu}-h^{\mu\nu}+O(h^{2}),$ 
but $\gamma_{\mu\nu}$ is commonly used for raising and lowering indices. It is thus not clear which semi-Riemannian manifold does this equality relate to. Such inconsistency can be ignored in the lowest order because it leads to the correct linear approximation of the wave equation. To extend the weak-field formalism to higher-order terms, the distinction between the premetric $\gamma$ and the average metric $q$ is necessary, as it exhibits the back-reaction of the waves on the background geometry. (Here and hereafter, indices of the metric tensors are sometimes suppressed for notational simplicity.)

Next step is to define the Ricci and Einstein tensors for an arbitrary metric $g$ as
\begin{eqnarray}\label{ricci}
\fl R_{\mu\nu}(g)\equiv [{\textstyle \frac{1}{2}}g^{\gamma\rho}
(g_{\rho\nu ,\mu}+
g_{\rho\mu ,\nu}-g_{\mu\nu ,\rho})]_{,\gamma}-
[{\textstyle \frac{1}{2}}g^{\gamma\rho}(g_{\rho\gamma ,\mu}+
g_{\rho\mu ,\gamma}-g_{\mu\gamma ,\rho})]_{,\nu}\nonumber\\ 
\lo{+}[{\textstyle \frac{1}{2}}g^{\gamma\delta}(g_{\rho\delta ,\gamma}+
g_{\rho\gamma ,\delta}-
g_{\gamma\delta ,\rho})][{\textstyle \frac{1}{2}}g^{\delta\rho}
(g_{\rho\nu ,\mu}+g_{\rho\mu ,\nu}-
g_{\mu\nu ,\rho})]\\
\lo{-}[{\textstyle \frac{1}{2}}g^{\gamma\rho}(g_{\rho\delta ,\nu}+
g_{\rho\nu ,\delta}-
g_{\nu\delta ,\rho})][{\textstyle \frac{1}{2}}g^{\delta\rho}
(g_{\rho\gamma ,\mu}+g_{\rho\mu ,\gamma}-
g_{\mu\gamma ,\rho})]\ ,\nonumber\\
\fl G_{\mu\nu}(g)\equiv R_{\mu\nu}(g)-{\textstyle \frac{1}{2}}g_{\mu\nu}g^{\alpha\beta}R_{\alpha\beta}(g)\ ,\nonumber
\end{eqnarray}
where $g^{\rho\tau}=(g)^{-1}_{\rho\tau}$, the same expressions apply to $\gamma$ and $q$. These equations remain a tensor even if we transfer to another semi-Riemann space (the reason is that covariant tensors are defined on a {\it metric} space rather than on some particular semi-Riemann one). From the proposals (i)-(iii) it follows that
$G_{\mu\nu}(\gamma)=0=G_{\mu\nu}(g)\,, G_{\mu\nu}(q)\neq 0$. 

Now, the {\it differences} between the covariant components of the above metrics are introduced,
\begin{eqnarray}
h_{\mu\nu}&\equiv &g_{\mu\nu}-q_{\mu\nu}\ ,\label{x1.72}\\
\eta_{\mu\nu}&\equiv &q_{\mu\nu}-\gamma_{\mu\nu}\nonumber\ .
\end{eqnarray}
It is necessary to specify the semi-Riemann space: for raising or lowering indices and for covariant differentiation the {\it averaged non-vacuum metric} $q$ will be used. Consequently, $h$ and $\eta$ are tensor fields on the semi-Riemann manifold $(M,q)$, i.e.
\begin{eqnarray}
h^{\mu\nu}&\equiv &q^{\mu\alpha}q^{\nu\beta}h_{\alpha\beta}\ ,\\
\eta^{\mu\nu}&\equiv &q^{\mu\alpha}q^{\nu\beta}\eta_{\alpha\beta}\ .\nonumber
\end{eqnarray}
Treating $h_{\mu\nu}$ as a perturbation of the metric $q_{\mu\nu}$  the Ricci tensor (\ref{ricci}) can be expanded in a power series
\begin{equation}\label{e3.5}
R_{\mu\nu}(g)=R_{\mu\nu}^{(0)}(q)+R_{\mu\nu}^{(1)}(q,h)+R^{(2)}_{\mu\nu}(q,h)+
R^{(3)}_{\mu\nu}(q,h)+O(h^{4})\ ,
\end{equation}
where
\begin{eqnarray}
\fl R^{(0)}_{\mu\nu}(q)\equiv R_{\mu\nu}(q)\ ,\nonumber \\ 
\fl R^{(1)}_{\mu\nu}(q ,h)\equiv {\textstyle \frac{1}{2}}q^{\rho\tau}\left(h_{\tau\mu ;\nu\rho}+h_{\tau\nu ;\mu\rho}
-h_{\rho\tau ;\mu\nu}-h_{\mu\nu ;\rho\tau}\right)\ ,\label{2.7}\\
\fl R^{(2)}_{\mu\nu}(q ,h)\equiv {\textstyle \frac{1}{2}}\left[{\textstyle \frac{1}{2}}h^{\rho\tau}{}_{;\mu}h_{\rho\tau ;\nu}+
h^{\rho\tau}\left(h_{\rho\tau ;\mu\nu}+h_{\mu\nu ;\rho\tau}-h_{\tau\mu ;\nu\rho}-
h_{\tau\nu ;\mu\rho}\right) \right. +\nonumber \\
\left. h^{\tau}{}_{\nu}{}^{;\rho}\left( h_{\tau\mu ;\rho}-h_{\rho\mu ;\tau}\right)
-\left(h^{\rho\tau}{}_{;\rho}-{\textstyle \frac{1}{2}}h^{\rho}_{\rho}{}^{;\tau}\right)\left(h_{\tau\mu ;\nu}+h_{\tau\nu ;\mu}-h_{\mu\nu ;\tau}
\right)\right]\; .\nonumber
\end{eqnarray}
Analogously,
\begin{equation}\label{e3.8}
R_{\mu\nu}(\gamma)=R_{\mu\nu}^{(0)}(q)+R_{\mu\nu}^{(1)}(q,(-\eta))+R^{(2)}_{\mu\nu}(q,(-\eta))+
O(\eta^{3})\ .
\end{equation}
It is obvious that ${R_{\mu\nu}^{(1)}(q,(-\eta))=-R^{(1)}_{\mu\nu}(q,\eta)}$ and ${R^{(2)}_{\mu\nu}(q,(-\eta))=R^{(2)}_{\mu\nu}(q,\eta)}$.
According to assumptions that both $g$ and $\gamma$ are vacuum metrics the following relation holds
\begin{eqnarray}
\fl 0&=&R_{\mu\nu}(g)-R_{\mu\nu}(\gamma)\nonumber\\
\fl  &=&R_{\mu\nu}^{(1)}(q,h)+R^{(2)}_{\mu\nu}(q,h)+R^{(1)}_{\mu\nu}(q,\eta)+
R^{(3)}_{\mu\nu}(q,h)+O(h^{4})+O(\eta^{2})\ .\label{e3.9}
\end{eqnarray}
At this point Efroimsky sets three assumptions:
\begin{enumerate}
\item[]
\hskip -0.5cm
{\bf Assumption 1.} {} The perturbations $h$ and $\eta$ are small in the sense that the terms of the orders $O(h^{4})$ and
$O(\eta^{2})$ are negligible.
\item[]
\hskip -0.5cm
{\bf Assumption 2.} {} The perturbations $\eta$ and $h^{2}$ are of the same order.
\item[]
\hskip -0.5cm
{\bf Assumption 3.} {} The tensor field $h$ consists of modes with short wavelengths which do not exceed the given maximal value $L$.
\end{enumerate}
A physical interpretation of the perturbations given by (\ref{x1.72}) is thus the following: $h_{\mu\nu}$
characterizes {\it measurable gravitational waves} whereas $\eta_{\mu\nu}$ is a {\it shift of the background geometry} from vacuum premetric $\gamma$ to nonvacuum effective average metric $q$ due to the presence of gravitational waves.
This enables us to interpret the equation (\ref{e3.9}) as the wave equation for perturbations $h$ on the background $q=\gamma+\eta$.
To make this wave equation applicable, one has to express $\eta$ in terms of $h$. Using the Brill-Hartle averaging procedure \cite{bh} over a spacetime volume of size $L$ for (\ref{e3.9}) (Efroimsky considers only space averaging but when the measurement lasts much longer than the period of waves one can employ spacetime average) we obtain
\begin{equation}\label{e3.11}
R^{(1)}_{\mu\nu}(q,\eta)=-\langle R^{(2)}_{\mu\nu}(q,h)\rangle_{L}\ .
\end{equation}
The averaging brackets on the left-hand side are omitted because the term contains only the modes with wavelength greater than $L$. It is thus clear from (\ref{e3.11})  and (\ref{2.7}) that the Assumption~2. is natural since the left-hand side is linear in $\eta$ whereas the right-hand side is quadratic in $h$.

Let us finally recall the  derivation of the stress-energy tensor of gravitational waves. By analogy with the Ricci tensor expansion (\ref{e3.5}) the Einstein tensor of the vacuum premetric $\gamma$ is represented as a series
\begin{equation}
0=G_{\mu\nu}(\gamma)=G_{\mu\nu}(q)+G^{(1)}_{\mu\nu}(q,(-\eta))+O(\eta^{2})\ ,
\end{equation}
and the effective stress-energy tensor of gravitational waves is defined as
\begin{equation}\label{x1.80}
G_{\mu\nu}(q)=8\pi T_{\mu\nu}^{(gw)}\equiv R^{(1)}_{\mu\nu}(q,\eta)-{\textstyle
\frac{1}{2}}
q_{\mu\nu}q^{\alpha\beta}R^{(1)}_{\alpha\beta}(q,\eta)\ .
\end{equation}
From (\ref{e3.11}) it follows (considering the Brill-Hartle averaging) that this tensor fully agrees with that of Isaacson \cite{isaac}.

The main advantage of the above Efroimsky's perturbation method is the possibility to consistently treat all low-frequency gravitational waves, and to explicitly derive effective stress-energy tensor (influencing the background) in this case. It can be extended to non-vacuum spacetimes with $T_{\mu\nu}$ of ideal fluid and/or with a possible cosmological constant $\Lambda$, see \cite{efroimsky,efroimsky2}. However, there are some problems concerning {\it high-frequency} gravitational waves which will now be discussed.

\section{Modification to include high-frequency waves}

In this section we first explicitly demonstrate that one can not consistently apply Efroimsky's treatment on Isaacson's high-frequency waves \cite{isaac} because the Assumption~2. is not fulfilled in such a case. Then we will present possible solution to this problem.

Let us start with observation that it is the nonvacuum background curved by the presence of gravitational waves
--- not the vacuum premetric $\gamma$ --- which is the basis of Isaacson's non-linear approach. 
Therefore,  the {\it nonvacuum average metric} $q$ is considered as the background on which high-frequency gravitational waves $h$ propagate.

We wish to use  the Efroimsky formalism in the high-frequency regime such that the tensor field $h$
contains high-frequency modes. We assume that they have short wavelengths $\lambda$, and a small amplitude ${h=O(\varepsilon)}$, where $\varepsilon=\lambda/S\ll 1$ is a small parameter because $\lambda\ll S$, $S$ denoting  a typical scale on which the background changes substantially. 

Let us emphasize that we follow here the same definition of the symbol $O(\varepsilon^n)$ as 
in \cite{isaac}, namely $f=O(\varepsilon^n)$ if there exists a constant $C>0$ such that
$|f|<C\varepsilon^n$ as $\varepsilon\to 0$. The quantity $f$ need not necessarily be proportional to
$\varepsilon^n$, it \emph{can be even smaller} than $C\varepsilon^n$ for $\varepsilon\to0$. 
Therefore, the assumption ${h=O(\varepsilon)}$ does \emph{not} automatically imply that 
${h\sim\varepsilon}$. The spectrum of possible high-frequency waves is thus not a priori restricted, 
it is only required that their amplitudes  fall to zero at least linearly with~$\varepsilon$, 
i.e. ${|h(\varepsilon)|<C\varepsilon}$.

Since we can consider ${S=O(1)}$ it follows that ${O(\varepsilon)=O(\lambda)}$ and 
${\partial h\sim h/\lambda=O(1)}$.  In accordance with Isaacson's approach  (note 
that the decomposition now reads $g=q+h$, instead of the notation $g=\gamma +h$
used in \cite{isaac}) we obtain the following orders of magnitude for the derivatives
of the background $q$ and the perturbation $h$:
\begin{equation}\label{x1.4}
\begin{array}{rclcrcl}
q_{\mu\nu}&=&O(1)\ ,&&h_{\mu\nu}&=&O(\varepsilon)\ ,\\
q_{\mu\nu ,\alpha}&=&O(1)\ ,&&h_{\mu\nu ,\alpha}&=&O(1)\ ,\\
q_{\mu\nu ,\alpha\beta}&=&O(1)\ ,&&h_{\mu\nu ,\alpha\beta}&=&O({\varepsilon}^{-1})\ .
\end{array}
\end{equation}
This results in the orders of magnitude of the terms in the Ricci tensor expansion (\ref{e3.5}), (\ref{2.7}) as
\begin{equation}\label{rexp}
R^{(0)}_{\mu\nu}=O(1),\quad
R^{(1)}_{\mu\nu}=O({\varepsilon}^{-1}),\quad
R^{(2)}_{\mu\nu}=O(1),\quad R^{(3)}_{\mu\nu}=O(\varepsilon).
\end{equation}
To apply the Efroimsky approach in this case we must consider the decomposition $q=\gamma+\eta$, where $\gamma$ is  the vacuum premetric and $\eta$ represents (in this case) {\it substantial} shift of the background geometry due to the presence of high-frequency gravitational waves $h$. We also introduce the scale $L$, such that $\lambda\ll L\ll S$. This enables us simultaneously to consider an averaging procedure in accordance with the Isaacson approach, and also to introduce a meaningful cut-off scale $L$ even if the wavelengths of high-frequency waves are not assumed to reach this value.

Of course, the geometry shift $\eta$ does not contain high-frequency perturbations. Considering the wave equation (\ref{e3.9}) and using the Brill-Hartle averaging over a spacetime volume $L$ to obtain the equation (\ref{e3.11}) we get in a conflict with the Assumption~2. which prescribes $O(\eta)=O(h^{2})$. Indeed, if $h=O(\varepsilon)$ there should be $\eta=O(\varepsilon^{2})$. But the right-hand side of (\ref{e3.11}) is now of the order $O(1)$, see (\ref{rexp}), and the same magnitude should also have the left-hand side. Since $\eta$
{\it does not contain} high-frequency waves, it is essential that $\eta=O(1)$. This is obviously in contradiction with both Assumptions~1. and 2. In fact, it disables any consistent perturbation expansions in the powers of~$\eta$.

Let us now suggest a modification of the Efroimsky formalism which will incorporate also the above case of a ``substantial" change of the background geometry due to the presence of high-frequency waves. Instead of the perturbation expansion (\ref{e3.8}) we consider a formal decomposition of the Ricci tensor of the premetric $\gamma=q-\eta$, namely
\begin{equation}\label{ricci1}
0=R_{\mu\nu}(\gamma)=R_{\mu\nu}(q)+\Delta R_{\mu\nu}(q,(-\eta))\ ,
\end{equation}
by which equation the expression $\Delta R_{\mu\nu}$ is {\it defined}.
Both terms on the right-hand side of (\ref{ricci1}) are of the same order $O(1)$. Moreover, the quantity $\Delta R_{\mu\nu}$ is conserved with respect to the background geometry $q$ which is easily seen from the equation (\ref{ricci1}) and the relation $(R_{\mu\nu}(q))^{;\nu}=0$ (the differentiation relates to the background metric $q$).

The question concerning the gauge invariance of $\Delta R_{\mu\nu}$ with respect to generalized gauge transformations has been recently analyzed in detail by Anderson \cite{anderson} in connection with possible definitions of the wave equation and stress-energy tensor for gravitational waves. Let us consider an arbitrary coordinate transformation of the type
\begin{equation}\label{transform}
\ol{x}^{\mu}=x^{\mu}+\xi^{\mu}\ ,
\end{equation}
that does not change the functional form of the background geometry $q$, i.e. ${\ol{q}(\ol{x})=q(\ol{x})}$
so that ${\gamma(x)\to\ol{\gamma}(\ol{x})=q(\ol{x})-\ol{\eta}(\ol{x})}$. 
Now, to prove the invariance of $\Delta R_{\mu\nu}$ we adopt (a slightly modified) Anderson's argumentation.
Performing the above coordinate transformation (\ref{transform}) 
of the Ricci tensor decomposition (\ref{ricci1}) we obtain
\begin{equation}\label{ricci2}
\ol{R}_{\mu\nu}(q(\ol{x}))+\ol{\Delta R}_{\mu\nu}(q(\ol{x}),(-\ol{\eta}(\ol{x})))=\ol{R}_{\mu\nu}(\ol{\gamma}(\ol{x}))=0\ .
\end{equation}
Here $\ol{R}_{\mu\nu}$ and $\ol{\Delta R}_{\mu\nu}$ are the same as $R_{\mu\nu}$ and $\Delta R_{\mu\nu}$, respectively, because the definition (\ref{ricci}) is maintained in any coordinates. Evaluating the relation (\ref{ricci2}) at $\ol{x}=x$ we thus get $R_{\mu\nu}(q(x))=-\Delta R_{\mu\nu}(q(x),(-\ol{\eta}(x)))$, and using (\ref{ricci1}) we obtain
\begin{equation}\label{invariance}
\Delta R_{\mu\nu}(q(x),(-\eta(x)))=\Delta R_{\mu\nu}(q(x),(-\ol{\eta}(x)))\ .
\end{equation}
 A generalized gauge transformation is defined in \cite{anderson} as a transformation in which the quantity $\ol{\eta}(x)$ is substituted for $\eta(x)$ into the tensor expressions of interest. This incorporates, as a particular case, the well-known infinitesimal gauge transformation,
\begin{equation}
\ol{\eta}_{\mu\nu}(x)=\eta_{\mu\nu}(x)+\xi_{\mu ;\nu}+\xi_{\nu ;\mu}\ ,
\end{equation}
where $\eta$,$\,\xi$ and their derivatives are small. Obviously, the equation (\ref{invariance}) expresses a generalized gauge invariance of $\Delta R_{\mu\nu}$.

After introducing the above decomposition (\ref{ricci1}) and demonstrating its invariance we can now present modification and generalization of the Efroimsky formalism. 
Replacing the term $R^{(1)}_{\mu\nu}(q,\eta)$ by $-\Delta R_{\mu\nu}(q,(-\eta))$ in equations (\ref{e3.9}), (\ref{e3.11}), (\ref{x1.80}), and omitting the terms $O(\eta^{2})$ we obtain relations
\begin{eqnarray}
\fl&& R_{\mu\nu}^{(1)}(q,h)+R^{(2)}_{\mu\nu}(q,h)-\Delta R_{\mu\nu}(q,(-\eta))+
R^{(3)}_{\mu\nu}(q,h)+O(h^{4})=0\ ,\label{me3.9}\\
\fl&&\Delta R_{\mu\nu}(q,(-\eta))=\langle R^{(2)}_{\mu\nu}(q,h)\rangle_{L}\ ,\label{me3.11}\\
\fl&& G_{\mu\nu}(q)=8\pi \tilde{T}_{\mu\nu}^{(gw)}\equiv-\Delta R_{\mu\nu}(q,(-\eta))+{\textstyle
\frac{1}{2}}
q_{\mu\nu}q^{\alpha\beta}\Delta R_{\alpha\beta}(q,(-\eta))\ .\label{mx1.80}
\end{eqnarray}
In case when gravitational waves do not have high-frequency modes it is still possible to employ the expansion of ${\>-\Delta R_{\mu\nu}(q,(-\eta))}$ in powers of $\eta$ and use its dominant term $R^{(1)}_{\mu\nu}(q,\eta)$ instead. Thus we recover Efroimsky's previous results, cf. (\ref{e3.9}), (\ref{e3.11}), (\ref{x1.80}).

In general, however, expressing $\eta$ in terms of $h$
from the equation (\ref{me3.11}) becomes an extremely difficult task because it is no longer a linear equation for $\eta$.
To overcome this problem we can use the equation (\ref{me3.11}) and substitute for $\Delta R_{\mu\nu}$ into the remaining equations (\ref{me3.9}) and (\ref{mx1.80}). We obtain the relations
\begin{eqnarray}
\fl R_{\mu\nu}^{(1)}(q,h)+R_{\mu\nu}^{(2)}(q,h)-\langle R_{\mu\nu}^{(2)}(q,h)\rangle_{L}
+R_{\mu\nu}^{(3)}(q,h)+O(h^{4})=0\ ,\label{waveeq}\\
\fl -G_{\mu\nu}(q)=\langle R_{\mu\nu}^{(2)}(q,h)\rangle_{L} -
{\textstyle
\frac{1}{2}}q_{\mu\nu}q^{\alpha\beta}\langle R_{\alpha\beta}^{(2)}(q,h)\rangle_{L}\equiv -8\pi T^{BH}_{\mu\nu}\ .\label{backreaction}
\end{eqnarray}
The equation (\ref{backreaction}) is obviously in perfect accordance with the Isaacson result \cite{isaac} which represents the background response to the presence of high-frequency gravitational waves, using the Brill-Hartle averaging to introduce the effective stress-energy tensor $T^{BH}_{\mu\nu}$ for high-frequency gravitational waves. The equation (\ref{waveeq}) is the wave equation for perturbations $h$ on the average metric $q$. In the highest order of high-frequency approximation this clearly reduces to $R^{(1)}_{\mu\nu}=0$ which also fully reproduces Isaacson's result. Additional terms in (\ref{waveeq}) can be used for study of nonlinear effects on the wave propagation.

Notice finally another interesting consequence of the equation (\ref{me3.11}) and the gauge invariance 
(\ref{invariance}) of $\Delta R_{\mu\nu}$. This directly guarantees gauge invariance of the stress-energy tensor $T^{BH}_{\mu\nu}$ defined in (\ref{backreaction}) (in the highest order). Proof of this property was presented already in the classic work \cite{isaac}, using however much more complicated method.

\section{Concluding remarks}
In our contribution we have compared the Efroimsky \cite{efroimsky,efroimsky2} and the Isaacson \cite{isaac} self-consistent perturbation schemes which describe propagation of weak gravitational waves  on a cosmological background. In both these approaches the background is influenced by the waves, i.e. the non-linear effects are taken into account. The classical Isaacson method applies to high-frequency waves. On the other hand, the Efroimsky formalism is applicable to low-frequency gravitational waves but does not admit the high-frequency limit. We have suggested a modification of the Efroimsky formalism by employing the gauge-invariant decomposition 
(\ref{ricci1}) of the Ricci tensor, introduced recently by Anderson \cite{anderson}. The resulting generalized system of equations (\ref{me3.9})-(\ref{mx1.80}) fully recovers the Efroimsky results in the absence of high-frequency modes, in the high-frequency limit it reproduces Isaacson's formulae.

Although we have considered here for simplicity  only  vacuum metrics $\gamma_{\mu\nu}$ and $g_{\mu\nu}$, possible generalization to non-vacuum spacetimes is straightforward. In fact, Efroimsky already generalized his formalism to spacetimes with ideal-fluid-like matter and a cosmological term \cite{efroimsky,efroimsky2}, in case of the Isaacson high-frequency approach this has been done recently in \cite{podolsky-svitek}.

\section*{Acknowledgements}
The work was supported in part by the grants GA\v{C}R 202/02/0735 and GAUK 166/2003
of the Czech Republic and the Charles University in Prague.

\end{document}